\documentclass[11pt]{article} 
\usepackage{graphicx,graphics,floatflt,amssymb,epsf,rotate} 
\usepackage{color}

\def\bar {\overline}

\def\delms {\Delta M_s}

\def\delmd {\Delta M_d}

\def\be {\begin{equation}}
\def\ee {\end{equation}}
\def\bea {\begin{eqnarray}}
\def\eea {\end{eqnarray}}

\def\beq{\begin{equation}}
\def\eeq{\end{equation}}
\def\barr{\begin{eqnarray}}
\def\earr{\end{eqnarray}}

\begin{document} 
\begin{flushright} 
SINP/TNP/2008/19,  
RECAPP-HRI-2008-012, 
DFTT-26/2008
\end{flushright} 
 
\vskip 30pt 
 
\begin{center} 

  {\Large \bf A simultaneous explanation of the large phase in $\mathbf
    {B_s-\bar B_s}$ mixing and $\mathbf {B \to \pi\pi/\pi K}$ puzzles in
    $\mathbf R$-parity violating supersymmetry} \\

  \vspace*{1cm} \renewcommand{\thefootnote}{\fnsymbol{footnote}} { {\sf Gautam
      Bhattacharyya${}^1$}, {\sf Kalyan Brata Chatterjee${}^{1}$}, {\sf
      Soumitra Nandi${}^{2,3}$}
  } \\
  \vspace{10pt} {\small ${}^{1)}$ {\em Saha Institute of Nuclear Physics,
      1/AF Bidhan Nagar, Kolkata 700064, India} \\
    ${}^{2)}$ {\em Regional Centre for Accelerator-based Particle Physics,
                Harish-Chandra Research Institute, \\Chhatnag Road, Jhusi,
               Allahabad 211019, India }\\
    ${}^{3)}$ {\em Dipartimento di Fisica Teorica, Univ. di Torino and INFN,
       Sezione di Torino, I-10125 Torino, Italy}}\\
    
\normalsize 
\end{center} 

\begin{abstract}  
  Recent data on $B$ meson mixings and decays are, in general, in accord with
  the standard model expectations, except showing a few hiccups: (i) a large
  phase in $B_s$ mixing, (ii) a significant difference ($> 3.5 \sigma$)
  between CP-asymmetries in $B^\pm \to \pi^0 K^\pm$ and $B_d \to \pi^\mp
  K^\pm$ channels, and (iii) a larger than expected branching ratio in $B_d
  \to \pi^0\pi^0$ channel. We show that selective baryon number violating
  Yukawa couplings in $R$-parity violating supersymmetry can reconcile all the
  measurements.

\vskip 5pt \noindent 
\texttt{PACS Nos:~ 12.60.-i, 13.20.He, 14.40.Nd } \\ 
\texttt{Key Words:~~$B$ Meson mixings and decays, $R$-parity violation}
\end{abstract}

\renewcommand{\thesection}{\Roman{section}} 
\setcounter{footnote}{0} 
\renewcommand{\thefootnote}{\arabic{footnote}}

\noindent {\bf Introduction:}~ There is still a possibility that by the time
we start analyzing the LHC data, some indirect evidence of new physics would
pop up from $B$ meson mixings and decays. So far, {\em most} of the
measurements in the $B$-factories are in reasonably good agreement with the
standard model (SM). In some cases, they are {\em not}, but in most such cases
the uncertainties plaguing the low energy hadronic phenomena prevent us from
making any substantial claim for new physics (NP).  But, rather than searching
for individual solutions for these discrepancies taken separately, if we seek
for a collective solution and observe that all or most of them can be
reconciled by a single NP dynamics, then that indeed deserves attention.
Here, we focus on three such anomalies, which we call puzzles, for each of
which a departure from the SM expectation is noticed with a reasonable
statistical significance:

($i$) {\em The $B_s$ mixing puzzle}:~ A model-independent test of new physics
contributing to $B_s$ mixing was performed with the following parametrization:
\begin{equation}
C_{B_s} e^{2i \phi_{B_s}} = \frac{A_s^{\rm SM} e^{-2i\beta_s} + 
A_s^{\rm NP} e^{2i (\phi_s^{\rm NP} - \beta_s)}}{A_s^{\rm SM} e^{-2i\beta_s}}
\, , 
\end{equation}
where $\beta_s \equiv {\rm arg}(-V_{ts}V^{\ast}_{tb} /V_{cs}V^{\ast}_{cb})$
has the value $0.018 \pm 0.001$ in the SM. UTfit has got two solutions 
\cite{utfit1}: 
\begin{eqnarray}
\phi_{B_s} ({\rm deg})  &=& -19.9 \pm 5.6 ~~,~~
A_s^{\rm NP} / A_s^{\rm SM} = 0.73 \pm 0.35 ~;\nonumber \\
\phi_{B_s} ({\rm deg})  &=& -68.2 \pm 4.9 ~~,~~
A_s^{\rm NP} / A_s^{\rm SM} = 1.87 \pm 0.06 ~. 
\label{utfit}
\end{eqnarray}
The SM expectation of $\phi_{B_s}$ is zero. But the above numbers show that
$\phi_{B_s}$ deviates from zero by more than 3.7$\sigma$ for the first
solution, while the second solution is significantly more distant from the SM
expectation\footnote{The UTfit collaboration have presented an updated
  estimate at ICHEP2008 (talk by M. Pierini): $\phi_{B_s} = (-19 \pm 7)^\circ
  \cup (-69 \pm 7)^\circ$, which shows a 2.6$\sigma$ discrepancy with the SM
  expectation. In any case, as long as this deviation from the SM value
  remains sizable, the numerical exercise leading to our conclusion holds. We
  thank D.~Tonelli of the CDF Collaboration for bringing this to our notice.}.
It should be noted that here the theoretical uncertainty is small, so a
statistically significant non-zero $\phi_{B_s}$ would constitute an
unambiguous NP signal. Combining the two UTfit solutions, the allowed range of
the mixing-induced CP-asymmetry in the $B_s$ system is given by $S_{\psi\phi}
\in [0.35, 0.89]$ at 95\% C.L. \cite{Buras:2008nn}, where $S_{\psi\phi} \equiv
\sin 2(|\beta_s| - \phi_{B_s})$.

($ii$) {\em The $\pi K$ puzzle}:~ The observed direct CP-asymmetries in the
$\pi K$ channel \cite{hfag1}, 
\begin{equation}
a_{\rm CP}(B_d \to \pi^{\mp}K^{\pm}) = -0.097 \pm 0.012 ~~,~~
a_{\rm CP}(B^{\pm}\to \pi^{0}K^{\pm}) = 0.050 \pm 0.025 ~~,
\end{equation}
imply that $\Delta{a_{\rm CP}} = a_{\rm CP}(B^{\pm}\to \pi^{0}K^{\pm}) -
a_{\rm CP}(B_d \to \pi^{\mp}K^{\pm}) = 0.14 \pm 0.029$ differs from the naive
SM expectation of zero at 4.7$\sigma$ level. In the QCD factorization
approach, $\Delta{a_{\rm CP}} = 0.025 \pm 0.015$, which differs from the
experimental value by 3.5$\sigma$ . This is quite reliable as most of the
model-dependent uncertainties cancel in the difference \cite{soni07}.

On the other hand, the following CP-conserving observables, as ratios of
branching ratios \cite{hfag1}
\begin{eqnarray}
\label{rnrc}
R_n &=& {1\over 2}{{\cal{BR}}[B^0_d\to \pi^- K^+] +
{\cal{BR}} [\bar{B^0_d}\to \pi^+ K^-]\over{\cal{BR}}[B^0_d\to \pi^0 K^0] + 
[\bar{B^0_d}\to \pi^0 \bar{K^0}]} = 1.0 \pm 0.07 \, , \\
R_c &=& {2}{{\cal{BR}}[B^+\to \pi^0 K^+] + 
{\cal{BR}}[B^-\to \pi^0 K^-]\over{\cal{BR}}[B^+\to \pi^+ K^0] + 
[{B^-}\to \pi^- \bar{K^0}]} = 1.10 \pm 0.07\, ,
\end{eqnarray}
are both in excellent agreement with the SM in which each of them is expected
to be unity. The `puzzle' seems to lie in the asymmetries.

($iii$) {\em The $\pi\pi$ puzzle}:~ The ratio 
\begin{eqnarray}
  R_{\pi\pi} = {2 {\cal{BR}}(B_d^0\to \pi^{0}\pi^{0}) \over 
{\cal{BR}}(B_d^0\to \pi^{\pm}\pi^{\mp})}=0.51 \pm 0.10,
\label{rpipi1}
\end{eqnarray}
is in conflict with the expected relation ${\cal{BR}}(B_d^0\to
\pi^{\pm}\pi^{\mp}) >> {\cal{BR}}(B_d^0\to \pi^0\pi^{0})$.  More specifically,
what is expected, based on different theoretical models (naive factorization
\cite{ali1}, PQCD \cite{nlopqcd1}, QCDF \cite{nloqcdf1}), is
${\cal{BR}}(B_d^0\to \pi^{0}\pi^{0}) \simeq {\cal{O}}(\lambda^2)~
{\cal{BR}}(B_d^0\to \pi^{\pm}\pi^{\mp})$, while what is observed is
${\cal{BR}}(B_d^0\to \pi^{0}\pi^{0}) \simeq {\cal{O}}(\lambda)~
{\cal{BR}}(B_d^0\to \pi^{\pm}\pi^{\mp})$.  On the other hand,
\begin{equation}
  R_{a} = {{\cal{BR}}(B_d^0\to \pi^{-}\pi^{+}) \over 
{\cal{BR}}(B^+\to \pi^{+}\pi^{0})}=0.93 \pm 0.09,
\label{rpipi2}
\end{equation}
is in good agreement with the SM.

It was shown in \cite{li1} that only a large color-suppressed tree amplitude,
with other amplitudes as expected in the SM, can explain the $\pi\pi$ and $\pi
K$ data, though such a large amplitude is hard to extract from short-distance
dynamics. We also note that large electroweak penguin (EWP) effects can
resolve the $\pi\pi$ and $\pi K$ puzzles \cite{ewp}, but such large EWP
contributions do not arise within the existing theoretical models. The option
of suppressing the $B^0 \to \pi^+\pi^-$ and enhancing $B^0 \to \pi^0\pi^0$
branching ratios by pumping up the charming penguins faces a serious obstacle
when confronted with the $\pi K$ data \cite{zhuang}. Again, the
next-to-leading order contributions in QCD factorization approaches
\cite{nloqcdf1} might jack up $B^0\to \pi^0\pi^ 0$ branching ratio but then
$B^0\to \rho^0\rho^0$ branching ratio goes out of control. Thus, a collective
explanation for all anomalies is hard to obtain.

To account for the large phase in $b\to s$ transition, several new physics
models have already been proposed \cite{newbs}. In this short paper, we show
that some selective $R$-parity (more specifically, baryon-number) violating
couplings can not only provide a large phase encountered in $B_s$-$\bar{B}_s$
mixing but can also explain the $\pi\pi$ and $\pi K$ riddles at the same time.

\noindent {\bf $R$-parity violating couplings:}~ R-parity is a discrete
symmetry defined as $R=(-1)^{3B+L+2S}$, where $B$, $L$, and $S$ are
respectively the baryon number, lepton number and spin of a particle. $R$
equals $1$ for all SM particles and $-1$ for all superparticles. Unlike in the
SM, conservations of $B$ and $L$ in supersymmetric models are rather {\em ad
  hoc}, not motivated by any deep underlying principle. However, such
couplings are highly constrained \cite{reviews}. Here, we concentrate on
explicitly broken $B$-violating part of $R$-parity violation (B-RPV)
only. These are contained in the superpotential,
\begin{equation}
{\cal W}=\frac{1}{2} \lambda''_{ijk} U^c_i D^c_j D^c_k \; ,
\label{superpot}
\end{equation}
where the antisymmetry in the last two indices implies $\lambda''_{ijk} = 
- \lambda''_{ikj}$. Our selection of B-RPV couplings is motivated through the
following chain of arguments: 

($i$)~ First, we take only those product couplings which contribute to
$B_s$-$\bar{B}_s$ and $B_d$-$\bar{B}_d$ mixings via one-loop box
diagrams. These are $\lambda''_{i13} \lambda^{''*}_{i12}$ and $\lambda''_{i23}
\lambda^{''*}_{i21}$ respectively, where $i$ corresponds to all the three
singlet up-type flavors.

($ii$)~ $\lambda''_{i13} \lambda^{''*}_{i12}$, for $i=2$, contributes {\em at
  tree level} to $b \to c\bar{c}s$ ($B_d\to J/\Psi K_S$). This is a golden
channel for $\sin 2 \beta$ measurement, yielding $\sin 2\beta = 0.681 \pm
0.025$ \cite{hfag1}, which is slightly lower than the SM fit $(\sin
2\beta)_{\rm fit} = 0.75 \pm 0.04$ \cite{soni08}\footnote{Using the recent
  lattice measurements of the hadronic matrix elements, $B_K$ and $\zeta_s$
  (see Eq.~(\ref{zeta})), the authors of \cite{soni08} have speculated a
  possible role of new physics to account for the difference between the
  fitted $\sin 2 \beta = 0.87 \pm 0.09$ (without $V_{ub}$ as input) and the
  measured value of $\sin 2 \beta$, which is about 2.1$\sigma$ lower than the
  fitted value.}. Now, for any $i$, $\lambda''_{i23} \lambda^{''*}_{i21}$ does
contaminate $\sin 2 \beta$ extraction any way by contributing to
$B_d$-$\bar{B}_d$ mixing through one-loop box graphs. But, nevertheless, we
refrain from using $\lambda''_{213} \lambda^{''*}_{212}$ to avoid any
overwhelming tree level new physics imposition on the `$\sin 2 \beta$ golden
channel'.

($iii$)~ For a simultaneous solution of the $\pi K$ puzzle, we expect to
generate a numerically meaningful contribution to $B^\pm \to K^\pm \pi^0$. The
corresponding quark level process $b\to s u \bar{u}$ is triggered by
$\lambda''_{i13} \lambda^{''*}_{i12}$ for $i=1$, but {\em not} for $i=3$. For
this reason, we consider $i=1$ only as far the combination $\lambda''_{i13}
\lambda^{''*}_{i12}$ is concerned. Regarding the other combination
$\lambda''_{i23} \lambda^{''*}_{i21}$, again we select the $i=1$ case as only
this choice leads to $b \to d u \bar{u}$ ($B \to \pi \pi$) at the tree level.

%It should be remembered that as long as the squark part of the $i$th flavor is
%considered as a propagator, it hardly matters which $i$ we take as far as the
%formulation part is concerned.

($iv$)~ Thus we are left with two combinations: $\lambda''_{113}
\lambda^{''*}_{112}$ and $\lambda''_{123} \lambda^{''*}_{121}$. These consist
of three independent couplings: $\lambda''_{113}$, $\lambda''_{112}$ and
$\lambda''_{123}$.  The strongest constraint on $\lambda''_{113}$ comes from
$n-\bar{n}$ oscillation: $\lambda''_{113}< 0.002-0.1$ for $m_{\tilde q}
<200-600$ GeV \cite{goity}. On the other hand, double nucleon decay into two
kaons puts the most stringent constraint: $\lambda''_{112} < 10^{-15}
R^{-5\over2}$ with $R = {\tilde{\lambda} \over (M_{\tilde g} M^4_{\tilde
    q})^{1\over5}}$, the ratio between the hadronic and supersymmetry breaking
scale. For $R \sim 10^{-3}$, the constraint is very strong: $\lambda''_{112}
\sim 10^{-7}$; while for $R \sim 10^{-6}$, it gets pretty relaxed:
$\lambda''_{112} \sim 1$. The upper bound on $\lambda''_{123}$ is 1.25 arising
from the requirement of perturbative unification.

\noindent {\bf B-RPV contributions to observables:}~ The product coupling
$\lambda''_{113} \lambda^{''*}_{112}$ triggers $b \to s$ transition, while 
$\lambda''_{123} \lambda^{"*}_{121}$ leads to $b \to d$ transition. We define:
\begin{equation}
\label{btoq}
h(b \to s) \equiv
\lambda''^{\ \ast}_{113} \lambda''_{112} \, , ~~~
h(b \to d) \equiv
\lambda''^{\ \ast}_{123} \lambda''_{121} \, .
\end{equation}
These combinations contribute to $B_q$--$\bar{B}_q$ ($q=d,s$) mixing via two
kinds of box diagrams, one with internal $d^c$ quark and $\tilde{u}^c$ squark
and the other with $u^c$ quark and $\tilde{q}^c$ squark. They are given by
($x_f = m_f^2/{\tilde{m}}^2$) \cite{dkn} 
\begin{equation}
M_{12(q)}^{\rm B-RPV}  =  \frac{h^2(b \to q)}{192 
\pi^2 M_{\tilde q_R}^2}  M_{B_q} \hat{\eta}_{B_q}
f_{B_q}^2 B_{B_q} 
\left(\tilde{S_0}(x_{u}) + \tilde{S_0}(x_d)\right) \, ,
\label{mgamma-s-rpv}
\end{equation}
where 
\begin{equation}
\tilde{S_0}(x) = \frac{1+x}{(1-x)^2} + \frac{2x\log x}{(1-x)^3} \, .
\label{s0tilde}
\end{equation}
Above, we have assumed the relevant squarks, $\tilde{u}_R$ and $\tilde{q}_R$,
to be mass degenerate, and we have denoted the common squark mass by
$\tilde{m}$.

The product coupling $h(b \to s)$ also contributes at {\em tree
  level} to non-leptonic $B$ decays like $b \to d \bar{d} s$ and $b \to u
\bar{u} s$, like $B^+ \to K^0 \pi^+, B^+ \to K^+ \pi^0, B_d \to K^0 \pi^0,
B_d\to K^{+}\pi^{-} ,B_s \to \phi \pi^0, B_s \to \pi^+ \pi^-, B_s \to K^+ K^-$
and their CP conjugate decays\footnote{Contributions from lepton-number
  violating $\lambda'$-type couplings to CP-asymmetry in $B^+ \to \pi^+ K$
  channel have been studied in \cite{Bhattacharyya:1999xh}. A similar study
  with $\lambda'$ couplings affecting $B \to X_s \gamma$ channel has been
  performed in \cite{Bhattacharyya:2000cd}. Note that the B-RPV couplings we
  have considered in this paper would contribute to $B \to X_s \gamma$ too,
  but it can be kept under control.}. Similarly, $h(b \to d)$
provides new {\em tree level} contribution to different $B \to \pi \pi$ decay
modes\footnote{Interplay between $B_d$-$\bar{B}_d$ and $B_d \to \pi^+ \pi^-$
  with $\lambda'$-type couplings was studied in
  \cite{Bhattacharyya:2002uv}.}. Thus, different decay rates receive different
amount of SM and B-RPV contributions, and the net amplitude in each case
amounts to their coherent sum\footnote{It should be noted that for simplicity
  of our analysis we have neglected the contributions arising from $R$-parity
  conserving sector in all these cases. The leading contributions from this
  sector to non-leptonic $B$ decays would come at one-loop order, whereas the
  B-RPV contributions in those decays would proceed at tree level. }.  The SM
amplitude is calculated in the naive factorization model
\cite{ali1}. Considering the uncertainties in any such calculation, we rely on
observables which are either the ratio of branching ratios or CP-asymmetries
(in $B\to \pi K$ modes).  For the direct CP-asymmetries to proceed we need a
sizable strong phase difference between the SM and the B-RPV amplitudes, which
may be generated from final state interaction and rescattering. Indeed, the
weak phases of the B-RPV couplings are free parameters.  For simplicity, we
have not considered the mixing between the B-RPV operators and the SM
operators between the scale $M_W$ and $m_b$. The dominant effect, which is
just a multiplicative renormalization of the B-RPV operator, can be taken into
account by interpreting the B-RPV couplings to be valid at the $m_b$ scale and
not at the $M_W$ scale (thus, one should be careful in using the constraints
on the couplings and in comparing different limits, though the numerical
differences are not expected to be significant).

\noindent {\bf Numerical inputs:}~
Unless otherwise mentioned, all numbers are taken from
\cite{hfag1}.
The measured values of the mass differences ($\Delta M_q$) are
\begin{equation}
\delmd = (0.507\pm 0.005)~{\rm ps}^{-1} \; , \quad \quad
\delms = (17.77\pm 0.10 ({\rm stat}) \pm 0.07 ({\rm syst})) ~{\rm ps}^{-1} \; .
\label{deltams}
\end{equation} 
We require $\sin 2\beta$ to lie between $0.75 \pm 0.04$ (the SM fit value with
$V_{ub}$ as input) and $0.681 \pm 0.025$ (measured from the golden channel
$B_d \to J/\Psi K_S$).

We also use the recent lattice values of the bag factors
\cite{lattice}
\begin{equation}
\label{zeta}
f_{B_s} \sqrt{B_{B_s}}= 281\pm 21~{\rm MeV} \; ,\quad \quad
\zeta_s = {f_{B_s} \sqrt{B_{B_s}}\over f_{B_d} \sqrt{B_{B_d}}} = 
1.20 \pm 0.06 \;,
\end{equation}
and the short distance factors
\begin{equation}
\eta_{B_d} = \eta_{B_s} =0.55 \; , \quad \quad
S_0(x_t) = 2.327 \pm 0.044 \; .
\end{equation}
The relevant CKM elements are \cite{utfit2}
\begin{equation}
|V_{td}|=8.54(28)\times 10^{-3} \; ,\quad \quad
|V_{ts}|=40.96(61)\times 10^{-3} \; ,\quad\quad
\gamma = (75 \pm 25)^{\circ}\;,
\end{equation}
while the other elements are taken to be fixed at their central values.

\noindent {\bf Results:}~ We proceed by making two assumptions or working
conditions:

($i$) The strong phase difference between the SM amplitude and the
corresponding BSM amplitude is the same irrespective of whether it is $b \to
s$ or $b \to d$ transition. This assumption relies on flavor SU(3) symmetry.

($ii$) In order to calculate the amplitudes for different non-leptonic decay
modes we have followed naive factorization approach and considered 10\%
uncertainty over the SM amplitudes to cover the different (model-dependent)
non-factorizable corrections. For $B_d\to \pi^0\pi^0$ mode we have taken this
uncertainty to be 20\%, since the SM branching ratio for this mode is $N_c$
sensitive \cite{ali1}.
 
There are five parameters which we like to constrain: the magnitude of two
product couplings ($|\lambda''^{\ \ast}_{123} \lambda''_{121}|$ and
$|\lambda''^{\ \ast}_{113} \lambda''_{112}|$), their weak phases ($\Phi_D
\equiv {\rm Arg}~\left(\lambda''^{\ \ast}_{123} \lambda''_{121}\right)$ and
$\Phi_S \equiv {\rm Arg}~\left(\lambda''^{\ \ast}_{113}
  \lambda''_{112}\right)$), and the common strong phase difference between the
NP and the SM amplitude ($\delta_S$). We vary all of them simultaneously, and
constrain them by requiring consistency with the observables $\Delta{a_{\rm
    CP}}$, $R_n$, $R_c$, $R_{\pi\pi}$, $R_a$, $\sin 2\beta$, $\Delta M_d$,
$\Delta M_s$ and $\phi_{B_s}$.  We also use $R = {\cal BR}(B^0\to
\pi^+\pi^-)/{\cal BR}(B^0\to \pi^+ K^-) = 0.259\pm 0.023$ \cite{hfag1} to
constrain those parameters. Our results are plotted in Fig.~1 and
Fig.~2. Throughout our analysis we have taken $\tilde{m} = 300$ GeV; a few
percent variation of it will not qualitatively alter our conclusions.  
%\vskip 10pt
\begin{figure}[htbp]
\begin{center}
\begin{tabular}{cc}
\resizebox{60mm}{!}{\includegraphics{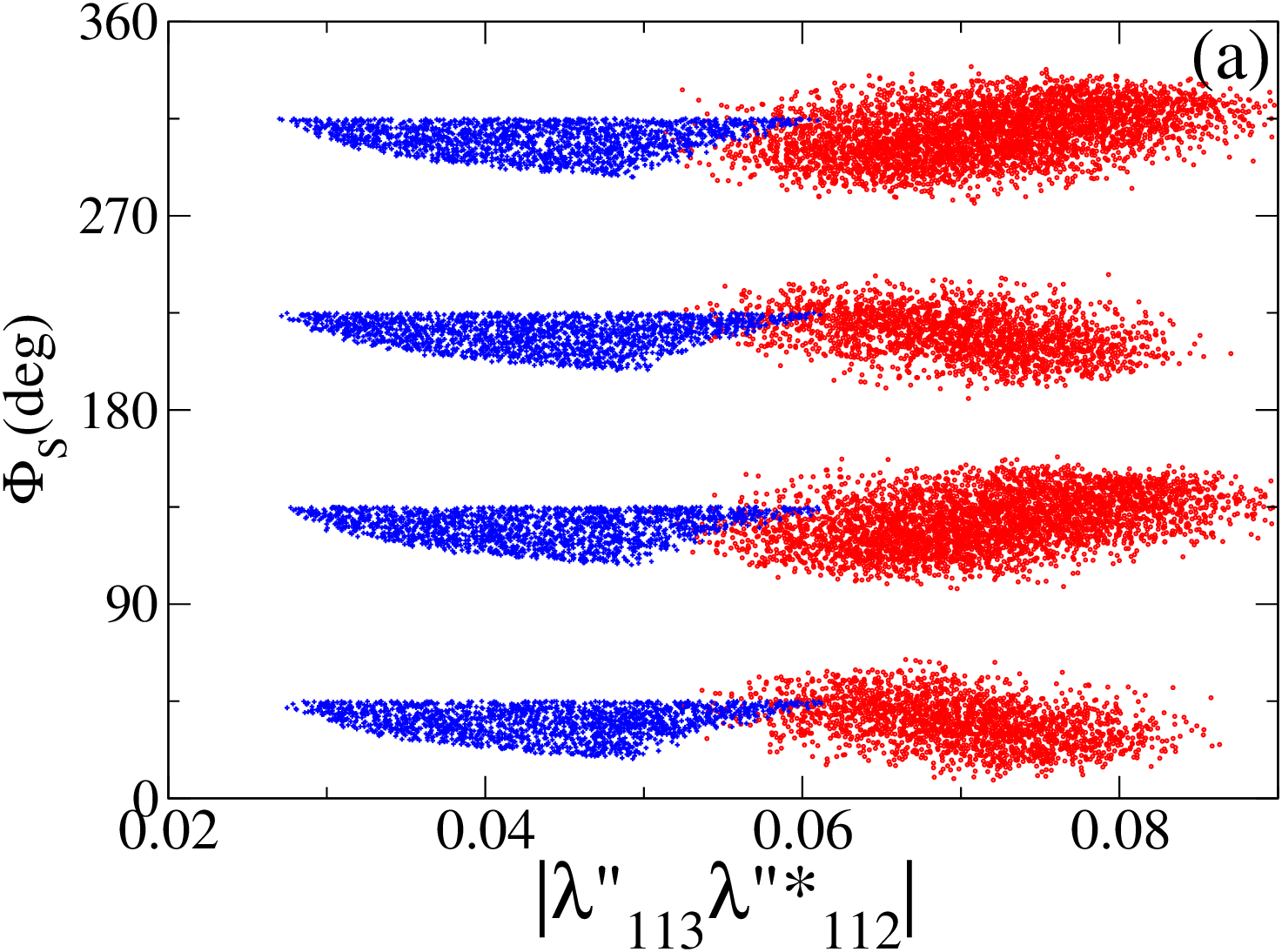}} &
\resizebox{60mm}{!}{\includegraphics{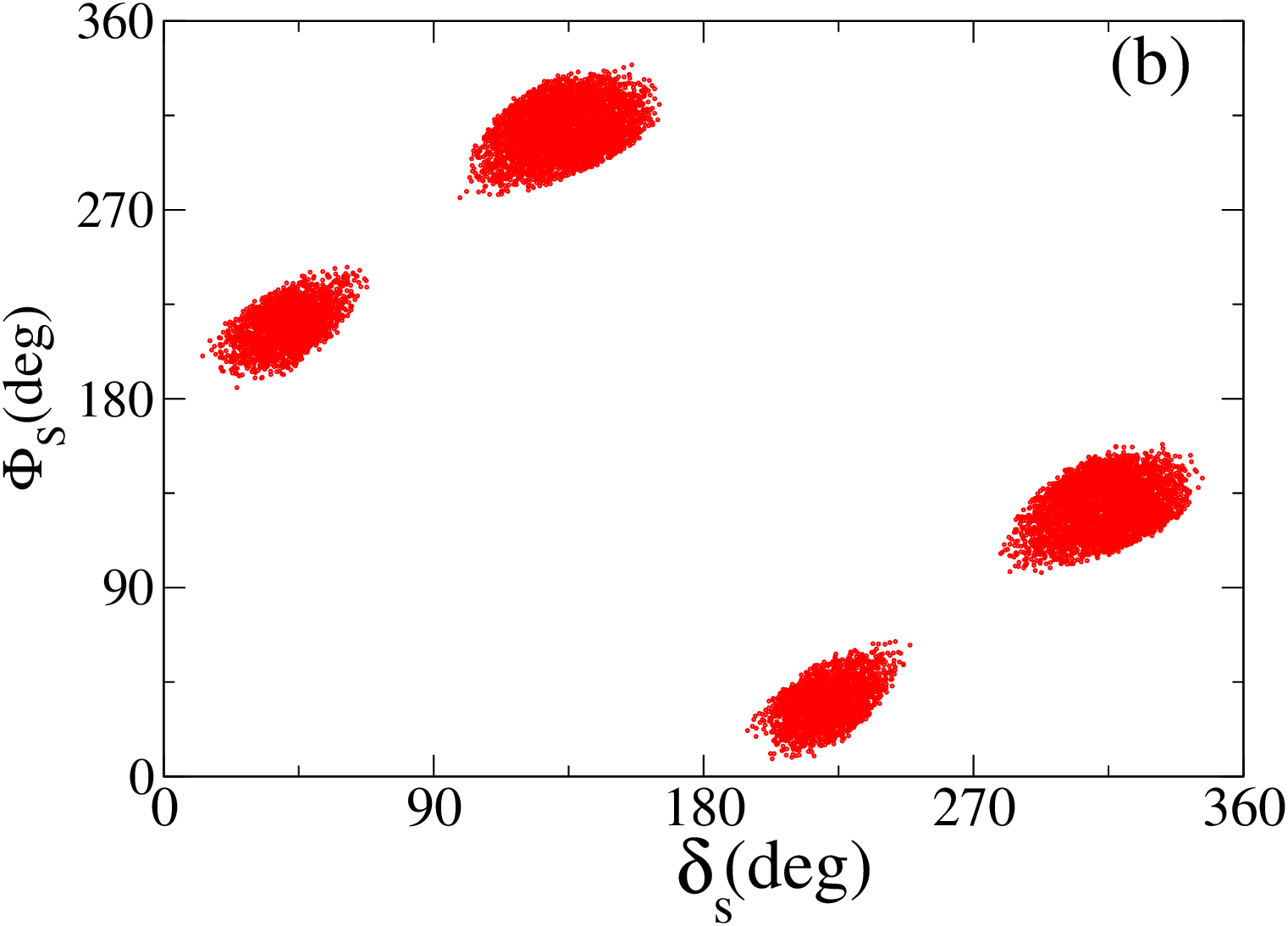}}\\
\end{tabular}
\caption{\small{\em{(Left panel-1a): The allowed zone in the plane of the
      magnitude of $h(b \to s)$ and its weak phase ($\Phi_S$) is shown. The
      red patches (on the right side) are scatter plots of the allowed
      parameters obtained by using $\Delta M_d$, $\sin2\beta$, $\Delta{a_{\rm
          CP}}$, $R$, $R_n$, $R_c$ and $R_a$; while the blue patches (on the
      left side) correspond to the space allowed by $\Delta M_s$ and
      $\phi_{B_s}$ only. (Right panel-1b): The allowed patches in the plane of
      the strong phase difference ($\delta_S$) and $\Phi_S$ are displayed.} }}
\label{Fig1}
\end{center}
\end{figure}
Although we varied all the parameters simultaneously, in Fig.~1a we projected
the allowed region in a two-dimensional space of the magnitude ($|\lambda''^{\
  \ast}_{113} \lambda''_{112}|$) and phase ($\Phi_S$) of $h (b \to s)$. The
red (right-side) patches are allowed solutions when all the five parameters
pass through the filters of $\Delta M_d$, $\sin2\beta$, $\Delta{a_{\rm CP}}$,
$R$, $R_n$ and $R_c$; while the blue (left-side) patches are zones allowed by
$\Delta M_s$ and $\phi_{B_s}$ only. There are {\em small} overlaps between the
allowed regions from the two sets. The overlaps signify a common solution for
all the three puzzles. With increasing statistics and with further reduction
in theoretical uncertainties, the overlap may increase or decrease, i.e. it
may or may not be possible to simultaneously address all the riddles with
B-RPV interactions. In Fig.~1b, we displayed the allowed zone in the plane of
$\Phi_S$ and $\delta_S$. We note at this stage that $\Phi_S$ has four sets of
solutions, one in each quadrant, and for each such set there is an associated
patch of $\delta_S$.

Note that $R_{\pi\pi}$ has been deliberately kept out of the above list of
constraints. If we include it, then to accommodate large ${\cal BR}(B_d^0\to
\pi^0\pi^0)$, only two sets of $\delta_S$ are allowed, one in the interval
$(100 \to 165)^{\circ}$ and the other in $(195 \to 245)^{\circ}$.  Since
$\delta_S$ has been {\em assumed} to be the common strong phase difference,
its limitations of the $b \to d$ sector infiltrate into the $b \to s$ sector
as well, thus eliminating $\Phi_S$ solutions in the second and the third
quadrants. The finally allowed values of $\Phi_S$ lie in the range $(10 \to
60)^{\circ}$ and $(275 \to 340)^{\circ}$. Clearly, if we relax the assumption
of {\em equality} of the strong phase difference (i.e. a {\em common}
$\delta_S$), $\Phi_S$ solutions in all the four regions will be allowed.

Fig.~2a is a zoomed version of Fig.~1a, except that in Fig.~2a we have
included {\em all} possible constraints at the same time. For illustration,
out of the two allowed sets of $\Phi_S$, the one within the range $(10 \to
60)^{\circ}$ has been shown. Fig.~2b is an equivalent description replacing
the magnitude and weak phase of $h (b \to s)$ by those of $h (b \to d)$. Note
that the constraint on $|h (b \to d)|$ is one order of magnitude tighter than
$|h (b \to s)|$, primarily because the SM prediction of the $B_d$ mixing is
relatively more precise.
%\vskip 10pt   
\begin{figure}[htbp]
\begin{center}
\begin{tabular}{cc}
\resizebox{60mm}{!}{\includegraphics{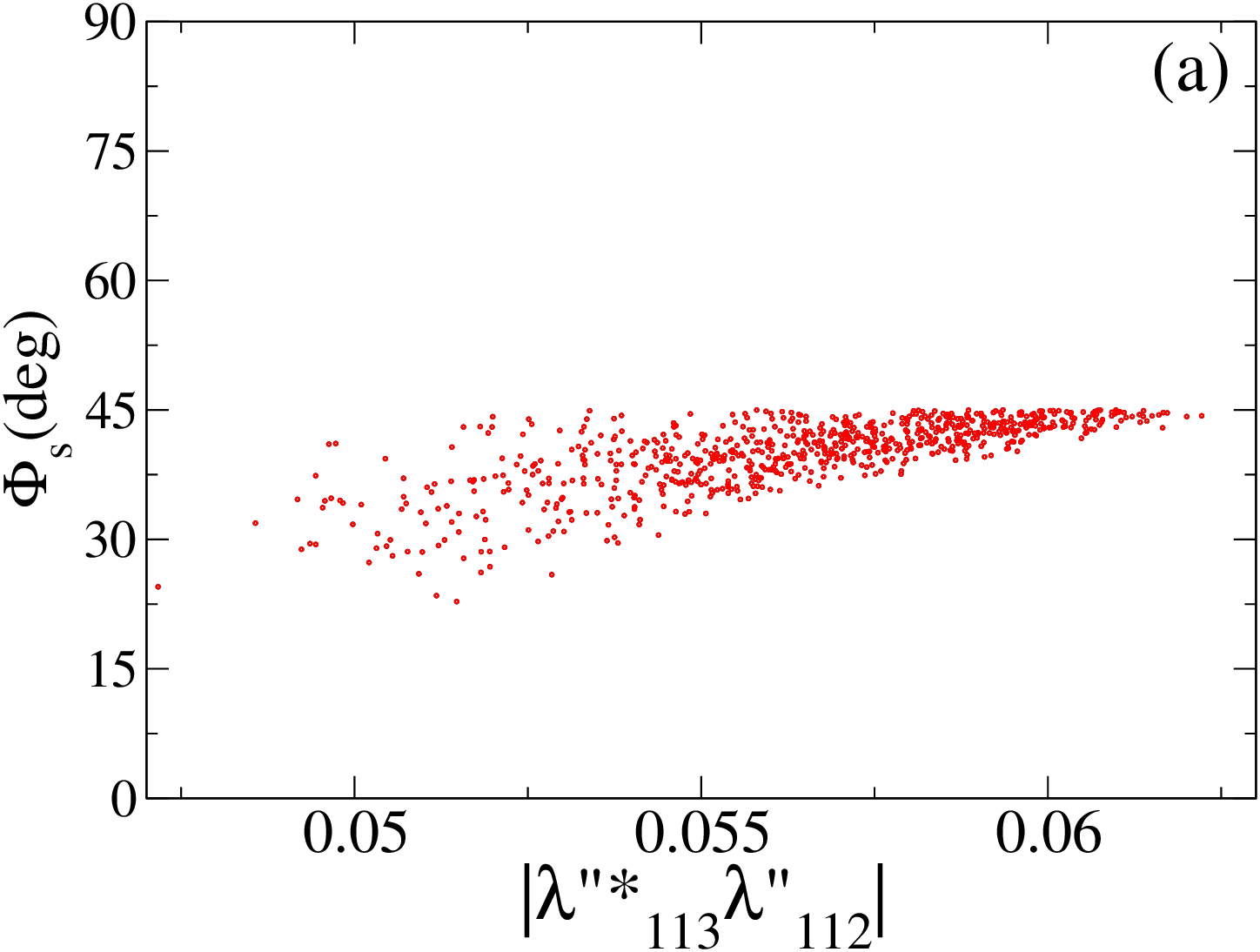}} &
\resizebox{60mm}{!}{\includegraphics{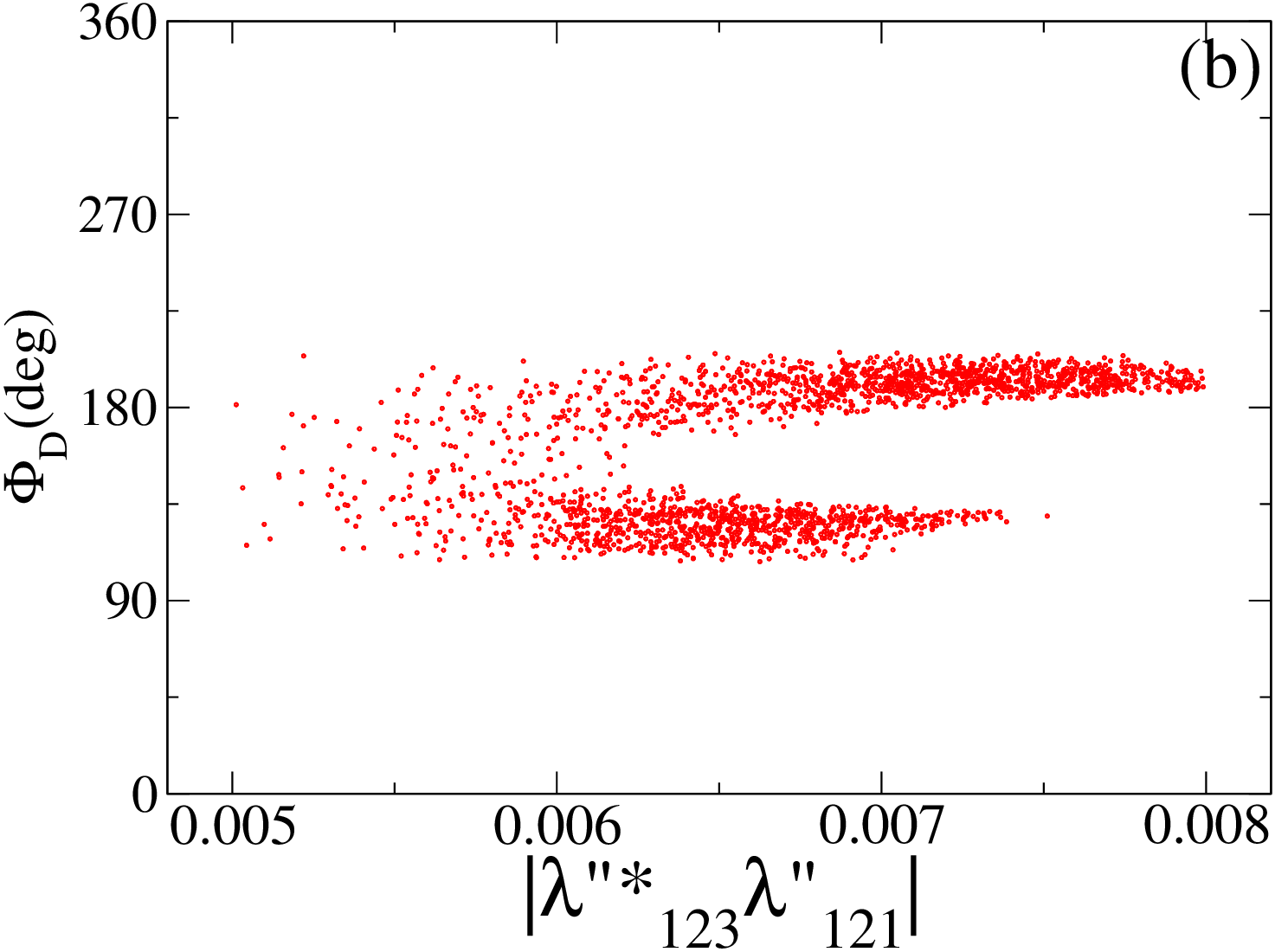}}\\
\end{tabular}
\caption{\small{\em{(Left panel-2a): Zoomed version of Fig.~1a, only that all
      constraints are now used, and focussed in the first quadrant solution
      of $\Phi_S$. (Right panel-2b): Similar to Fig.~2a, but in the space of
      the magnitude and phase of $h (b \to d)$.}}}
\label{Fig2}
\end{center}
\end{figure}

\noindent {\bf Conclusions:}~ In this paper, we wanted to solve three puzzles
in $B$ physics, namely, the large phase in $B_s$ mixing, a more than 3.5
$\sigma$ discrepancy between CP-asymmetries in charged and neutral $B$ decays
in $\pi K$ modes, and a significantly larger than expected neutral $B$ decay
in $\pi^0 \pi^0$ channel. Here we make two remarks: ($i$) the theoretical
uncertainty in the estimation of the $B_s$ mixing phase is small and hence a
large non-zero phase would constitute a clinching signal for new physics;
($ii$) but, on account of large hadronic uncertainties associated with the
$\pi K$ and $\pi \pi$ modes, the {\em discrepancies} observed in $\Delta
a_{\rm CP}$ and $R_{\pi\pi}$, though tantalizing, are not conclusive.  In
fact, to get rid of these theoretical uncertainties as much as possible, we
considered the {\em difference} between CP-asymmetries and the {\em relative}
branching ratios. Yet, from a conservative point of view, instead of entering
into a debate whether the discrepancies constitute `puzzles' or `non-puzzles',
all that we wanted to emphasize in this paper is that if one can figure out a
new dynamics beyond the SM that causes a simultaneous and systematic movement
of all those theoretical estimates towards better consistency with
experimental data, then that source of new physics calls for special
attention. As an illustration, we advanced the case of explicit baryon-number
violating part of supersymmetry, and we have used only two product couplings,
constructed out of three individual ones, to explain all the data.  One should
keep track of it in the LHC data analysis, as such interactions would give
lots of final state jets.

In fact, even within the $B$ physics context, it may be possible to infer our
choices of B-RPV couplings (or, similar type diquark couplings) from the
following observations: the coupling $h(b \to s)$ will contaminate $B_s \to
K^+ K^-$ ($b \to s u \bar{u}$ at the quark level) which is used to extract
$\gamma = {\rm Arg}~(V^\ast_{ub})$ \cite{Fleischer:1996aj}, but it would not
affect $B_s \to D_s K$ ($b \to s c \bar{u}$ at the quark level) which is also
used to determine $\gamma$ \cite{Aleksan:1991nh}. Any statistically different
measurement of $\gamma$ between these two methods will strengthen our
hypothesis. Moreover, either of the two methods would yield $\gamma$ different
from the value extracted from $B \to \pi K$.  We stress again that the
falsifiability of our hypothesis, under the assumptions spelt above, can be
judged from Fig.~1a by noting that the common solution zone in the parameter
space arising from the `$B_s$-set' and the other data set may shrink or expand
as more data accumulate. LHCb will definitely shed more light to these issues.

\noindent {\bf Acknowledgements:}~ GB acknowledges a partial support through
the project No.~2007/37/9/BRNS of BRNS (DAE), India. KBC acknowledges
hospitality at the Regional Centre for Accelerator-based Particle Physics of 
HRI, Allahabad, during part of his work. SN's work is supported in part by 
MIUR under contact 2004021808-009 and by a European Community's Marie-Curie 
Research Training Network under contract MRTN-CT-2006-035505 ``Tools and 
Precision Calculations for Physics Discoveries at Colliders''. We thank D. 
Tonelli and A. Kundu for their helpful comments on the manuscript.

\end{document}